\documentclass[useAMS,usenatbib]{mn2e}

\usepackage{graphicx}
\usepackage{epstopdf}
\usepackage{amssymb}
\usepackage{color}
\newcommand{\exosatme}{{\it EXOSAT}/ME}

\newcommand{\rxte}{{\it RXTE}}
\newcommand{\rxtepca}{{\it RXTE}/PCA}
\newcommand{\rxteasm}{{\it RXTE}/ASM}
\newcommand{\swiftbat}{{\it Swift}/BAT}
\newcommand{\cgrobatse}{{\it CGRO}/BATSE}
\newcommand{\ftools}{{\tt FTOOLS\/ v6.9}}
\newcommand{\isis}{{\tt ISIS}}
\newcommand{\sitar}{{\tt SITAR}}

\title[mHz QPOs in Cyg X-3]{The Reoccurrence of mHz QPOs in Cygnus X-3}
\author[K. I. I. Koljonen, D. C. Hannikainen and M. L. McCollough]
{K. I. I. Koljonen$^{1}$\thanks{E-mail: karri@kurp.hut.fi (KIIK); diana@kurp.hut.fi (DCH); mccolml@head.cfa.harvard.edu (MLM)}
D. C. Hannikainen$^{2,1,3}$ and M. L. McCollough$^{4}$\\
$^{1}$Aalto University Mets\"ahovi Radio Observatory,  Mets\"ahovintie 114, FI-02540 Kylm\"al\"a, Finland\\
$^{2}$Finnish Centre for Astronomy with ESO (FINCA), University of Turku, V\"ais\"al\"antie 20, FI-21500 Piikki\"o, Finland\\
$^{3}$Department of Physics and Space Sciences, Florida Institute of Technology, 150 W. University Blvd., Melbourne, FL 32901, USA\\
$^{4}$Smithsonian Astrophysical Observatory, 60 Garden Street, Cambridge, MA 02138-1516, USA}

\begin{document}

\date{Accepted . Received ; in original form }

\pagerange{\pageref{firstpage}--\pageref{lastpage}} \pubyear{ }

\maketitle

\label{firstpage}

\begin{abstract}

We have re-analyzed archival \rxte\/ data of the X-ray binary Cygnus X-3 with a view to investigate the timing properties of the source. As compared to previous studies, we use an extensive sample of observations that include all the radio/X-ray spectral states that have been categorized in the source recently. In this study we identify two additional instances of Quasi-Periodic Oscillations that have centroid frequencies in the mHz regime. These events are all associated to a certain extent with major radio flaring, that in turn is associated with relativistic jet ejection events. We review briefly scenarios whereby the Quasi-Periodic Oscillations may arise. 

\end{abstract}

\begin{keywords}
Accretion, accretion discs -- Binaries: close -- Stars: oscillations -- X-rays: binaries -- X-rays: individual: Cygnus X-3 -- X-rays: stars
\end{keywords}

\section{Introduction}

Cygnus X-3 is a well-known X-ray binary (XRB) located in the plane of the Galaxy at a distance of approximately 9 kpc \citep{dickey,predehl}. Its discovery dates back to 1966 \citep{giacconi} but the intrinsic nature of the system remains a mystery despite extensive multiwavelength observations throughout the years. A truly puzzling feature of Cygnus X-3 is that the 4.8-hour periodicity in the X-ray lightcurve \citep{parsignault} attributed to orbital modulation is characteristic of typical low-mass XRBs, but infrared observations suggest the mass-donating companion to be a Wolf-Rayet (WR) star \citep{keerkwijk}, characteristic of high-mass XRBs.

Unlike most other XRBs, Cygnus X-3 is relatively bright in the radio virtually all of the time ($\gtrsim$ 100 mJy), with periods of quenched or quiescent radio emission punctuated by giant outbursts of the type unseen in classical radio-emitting XRBs \citep{waltman}. During these outbursts there is strong evidence of relativistic jet-like structures \citep{mio,millerjones}. 

Surprisingly, given the wealth of available timing data, the X-ray timing properties of Cygnus X-3 are not particularly well studied. Thus, apart from the strong orbital modulation present in the X-ray lightcurve, the timing properties of Cygnus X-3 appear remarkably nondescript. The power density spectrum (PDS) of Cygnus X-3 was studied in \citet{willingale} and was found to be well described by a power law of index $\beta=-1.8$ in the frequency range 10$^{-5}$--0.1 Hz. This result has been more or less verified in further studies by Choudhury et al. (2004; $\beta \sim -1.5$) and Axelsson et al. (2009; $\beta \sim -2$ in the canonical, disk-dominated soft X-ray state and $-1.8$ in the canonical, Comptonization-dominated hard X-ray state). The PDS has no power above 0.1 Hz, a result which was discussed in Berger \& van der Klis (1994; they placed an upper limit of 12\% rms above 1 Hz) and subsequently in \citet{axelsson}. This is believed to be an effect of scattering in the nearby surrounding medium \citep{zdziarski}. However, \citet{vanderklis} found short intervals (5--40 cycles) of mHz quasi-periodic features in the PDS during the soft state and possibly also during the hard state in \exosatme\/ data. A mHz feature in the Cygnus X-3 hard X-ray lightcurve was also found in a balloon flight study by \citet{rao}. However, no Quasi-Periodic Oscillations (QPO) were found in the \rxtepca\/ study in \citet{axelsson} in either the soft or hard spectral states of the source. It is
important to note that all of these previously studied PDSs are indicative of red noise (the ``leakage'' of power in low frequency features to higher frequencies) dominated PDSs. The discrepancies between the reported results clearly show that these transient features are short-lived and sporadic or time-sensitive events. Interestingly, the previously discovered QPOs appear to occur following major radio flaring episodes \citep{koljonen2}.

In this paper we study the PDS of Cygnus X-3 using the extensive archive of \rxtepca\/ timing data including pointings during and after major radio flaring events. We introduce the data and the analysis method in Section 2 and present our results in Section 3. In Section 4 we briefly review possible causes for the QPOs in Cygnus X-3 and summarise the paper in Section 5.

\section[]{Observations and method for current study}

\subsection{Observations}

We analyzed, following standard procedures using \ftools\/, the archived \rxte\/ observations from 1997 to 2011 totaling 172 pointings and more than 0.7 Msec of data, covering all spectral states as defined in Koljonen et al. (2010; hereafter K10), and orbital phases. The \rxtepca\/ lightcurves were extracted using the generic binned data with a time resolution of 4 ms from channels 0--35 corresponding approximately to the energy range 2--15 keV (epoch dependent). Subsequently, the lightcurves were binned to a 100 ms resolution. We verified our results by comparing the outcome to Standard 1 data that have a time resolution of 125 ms. In order to gauge the X-ray spectral state\footnote{Following K10 from hard to soft: quiescent, transition, flaring/hard X-ray (FHXR), flaring/intermediate (FIM), flaring/soft X-ray (FSXR) and hypersoft state.} of the source during the selected pointings (if not already done so in K10),  we extracted lightcurves from channels 3--11 and 24--37 corresponding to energy ranges 3--6 keV and 10--15 keV using the Standard 2 data products and compared the resulting hardnesses and intensities of these bands to the hardness-intensity diagram presented in K10.

Furthermore, in order to identify the correct radio/X-ray state simultaneous radio observations were inspected using archival data from the Green Bank Interferometer (GBI), Ryle/AMI-LA, and RATAN-600. To follow the overall evolution of the radio/X-ray spectral states within the context of the selected pointings containing the QPOs, the monitoring data from \rxteasm\/, \swiftbat\/ and \cgrobatse\/ were used in phase-selected format as in K10 in addition to the radio lightcurves from the above-mentioned radio observatories.

\subsection{Timing analysis}

The timing analysis in this study was undertaken using  \isis\/ (Interactive Spectral Interpretation System, \citealt{houck}) with the \sitar\/ timing analysis module. The usual procedure for searching for quasi-periodic phenomena in lightcurves is to construct a power density spectrum (PDS) and search for sharp peaks in this spectrum. To construct a PDS we prepared the 100 ms time resolution lightcurves (done separately for each pointing) and calculated the PDS for segments of length 8192 bins over the whole lightcurve rejecting the segments with data gaps and finally averaging the PDS over the whole lightcurve. To determine the significance level of the possible QPO detections we followed the Monte-Carlo analysis of \citet{benlloch} which relies on simulating random red-noise dominated lightcurves with the algorithm described in \citet{timmer}. First, we performed simple fits to the PDS with \isis\/ (with high degree of binning), where we include a simple powerlaw for modelling the low-frequency noise and a constant for modelling the Poisson noise. Then 5000 lightcurves were simulated, using the model of the PDS ensuring that the generated lightcurves had the same length and mean counts as the original observations. Based on the distribution of the resulting PDSs from simulated lightcurves for each frequency bin, 95\%, 99\% and 99.9\% confidence intervals were obtained. Also, root-mean square values were obtained from the dynamically calculated PDSs from a frequency range 0.004--0.1 Hz using the same segment length over the lightcurves as above. 

The orbital phase of individual segments of lightcurves were determined using a cubic ephemeris \citep{singh}.

\section[]{Results}

\subsection[]{Previous detections of QPOs}

\begin{table*}
\centering
\caption{Log of QPOs identified from Cyg X-3 in previous studies.}
\begin{tabular}{ccllcclcc}
\hline
Mission$^{a}$ & ObsID & \multicolumn{2}{c}{Date} & Phase & TSMF$^{b}$ & Centroid freq. & X-ray state \\
&  &  & (MJD) & & (days) & (mHz) \\
\hline
\hline
E & & 1983 Oct 22 & 45629.11--45629.18 & 0.41--0.75 & 11.5 & 2.6 & high \\
E & & 1983 Nov 5 & 45643.01--45643.15 & 0.14--0.54 & 25.5 & 2/15 & high \\
E & & 1983 Dec 18 & 45686.14--45686.22 & 0.13--0.37 & 68.6 & 1.7 & high \\
E & & 1984 Jan 3 & 45702.55--45602.73 & 0.18--0.64 & 84.9 & 0.7/3.4 & high \\
E & & 1984 May 21 & 45841.64--45841.92 & 0.00--0.13 & -- & 1 & low \\
\hline
B & & 1986 Mar 18 & 46507.1 & -- & -- & 8 &  ---  \\
\hline
\hline
\label{table-1}
\end{tabular}
\begin{list}{}{}
\item[$^{\mathrm{a}}$] E=EXOSAT (1--10 keV); B=balloon flight (20--100 keV).
\item[$^{\mathrm{b}}$] Time Since Major Flare. ``$-$" corresponds to not known.
\end{list}
\end{table*}

QPOs have been detected at least six times according to the literature in the first studies dedicated to this topic (Table 1). \exosatme\/ observations exhibited QPOs with amplitudes between 5 and 20\% of the 1--10 keV flux and periods in the range 0.7--15 mHz (50--1500 s) which persisted for 5--40 cycles and occurred in the phase interval 0.0--0.75 \citep{vanderklis}. A balloon study \citep{rao} showed a QPO in the 20--100 keV lightcurve with a frequency of 8 mHz (121 s), pulse fraction 40\% and duty cycle 60\% during maximum brightness.

A later study based on \rxte\/ pointing data from 1996 through 2000 was conducted by \cite{axelsson} and, as mentioned above, they found no evidence of QPOs on any timescale. Specifically, on the shorter timescales ($>10^{-3}$ Hz) they do not detect QPOs at all and the power density spectrum is well-described by a powerlaw of index $-$2, while on longer timescales ($<10^{-3}$ Hz), they find that the variability is dominated by the state transitions. However, as \cite{scott} point out in their study of timing noise in the Crab, random walk processes will yield a powerlaw index of $-$2. This suggests that the flaring in Cygnus X-3 produces the red noise that essentially mimics a random walk process and thus hides any potential intrinsic variability.

\subsection[]{Current study}
\begin{figure}
\begin{center}
  \includegraphics[width=0.42\textwidth]{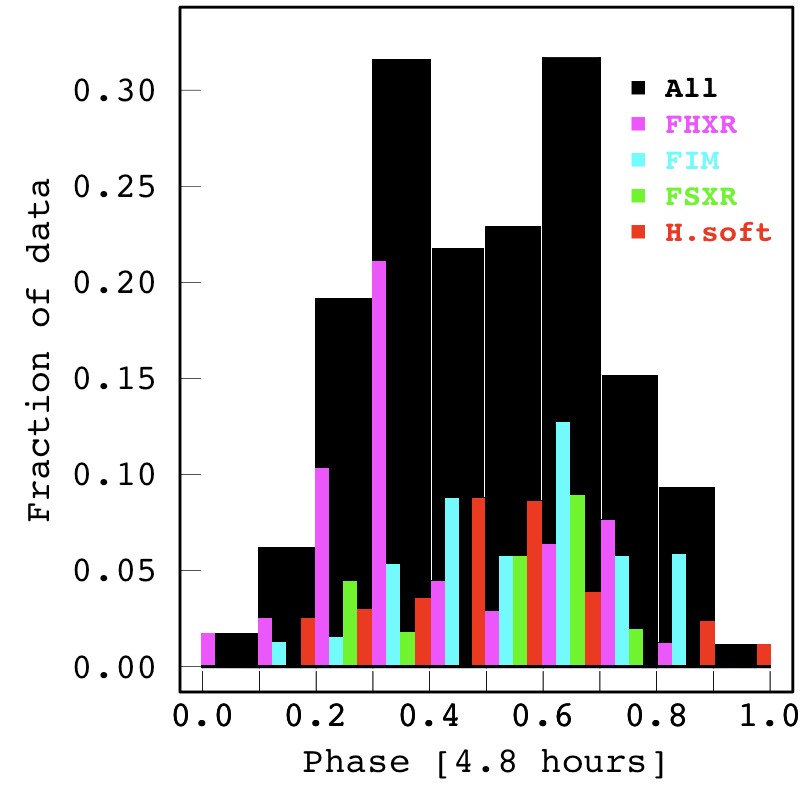}
\caption{The fraction of data whose root-mean square values are above the level of a standard deviation $+$ the median rms of all the data in a bin spanning a tenth of a phase (black histogram). The coloured histograms show how the fractions of data are divided among states.}  
  \label{fig-1}
\end{center}
\end{figure}
\begin{figure}
\begin{center}
  \includegraphics[width=0.5\textwidth]{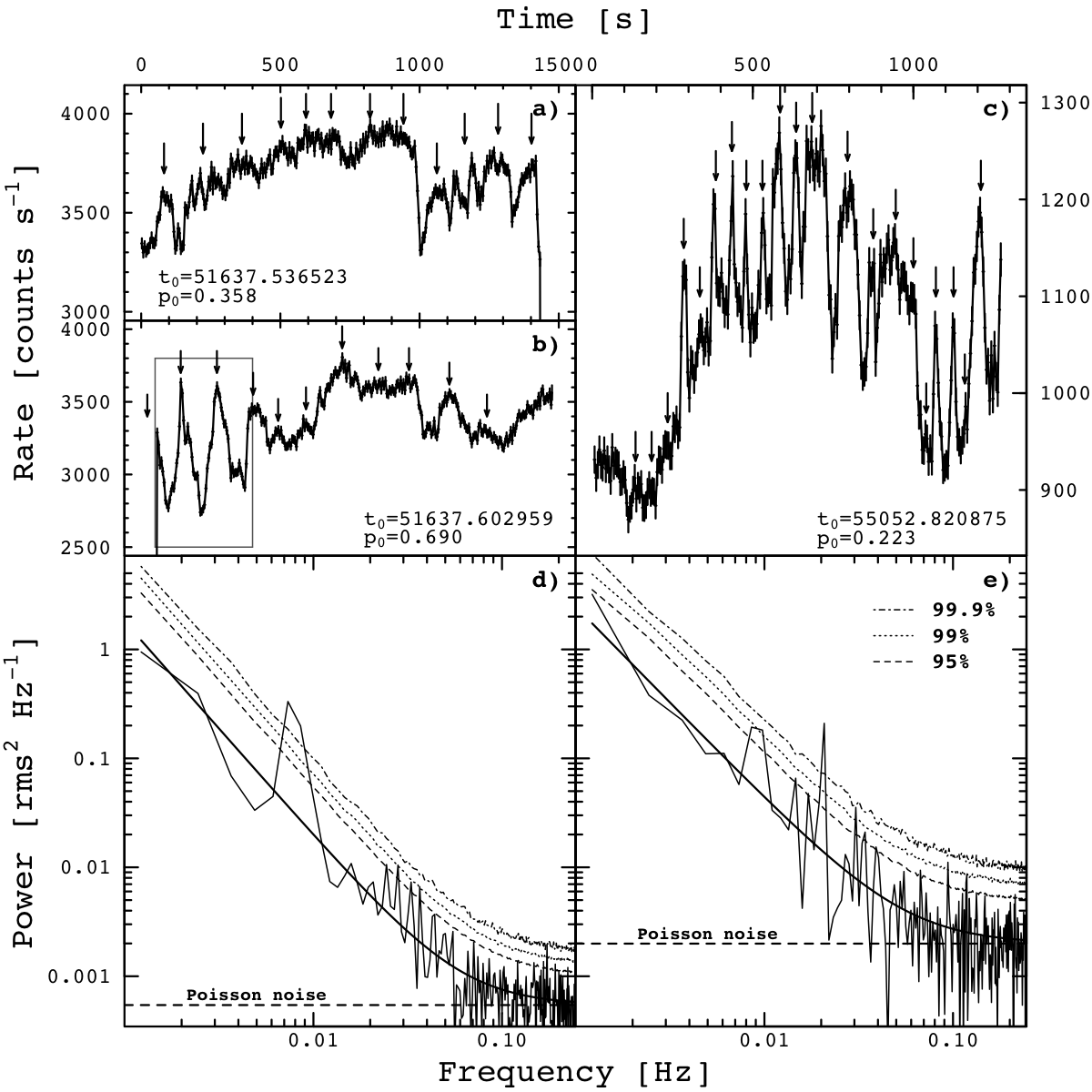}
\caption{The panels show the \rxtepca\/ PDSs from each pointing tabulated in Table \ref{table-2}. Panels \textit{a} and \textit{b} show the lightcurves from the 2000 Apr 3 pointing and panel \textit{c} for the 2009 Aug 9 pointing (binned in 5 sec bins, t$_{0}$ marks the MJD and p$_{0}$ the phase for the first bin) with arrows marking the QPOs. The box in panel \textit{b} shows the area that is zoomed in Figure~\ref{fig-3}. Panel \textit{d} shows the PDS from the 2000 Apr 3 pointing and panel \textit{e} from the 2009 Aug 9 pointing with Monte-Carlo significance and Poisson noise levels. See text for details on methods.}   
  \label{fig-2}
\end{center}
\end{figure}
\begin{figure}
\begin{center}
 \includegraphics[width=0.42\textwidth]{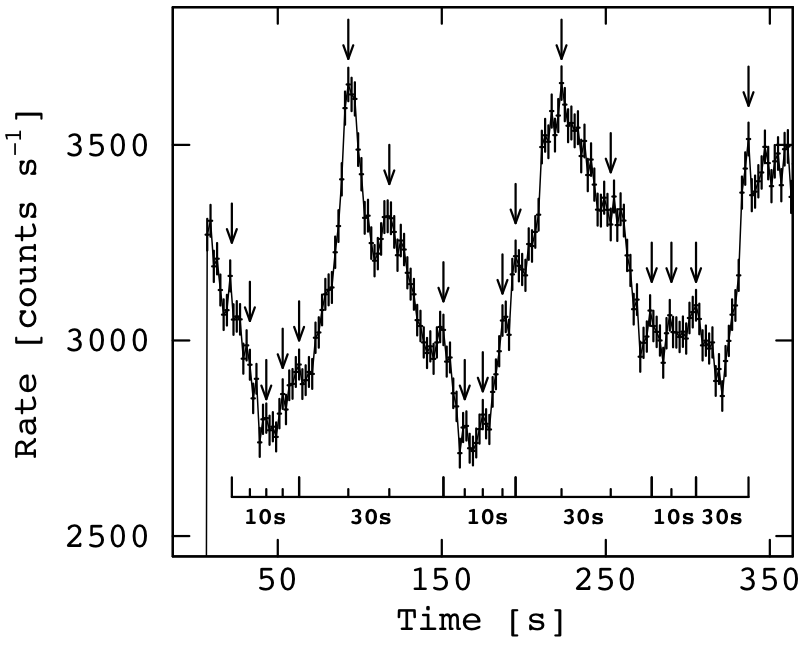} \\
\caption{A blow-up of the beginning of the lightcurve in Fig. \ref{fig-2}\textit{b} showing strong pulsations of $\sim$ 120 sec and superposed on that two weaker sequences of 30 sec and 10 sec (marked with arrows) in alternating fashion.}
  \label{fig-3}
\end{center}
\end{figure}

Our study yielded \textit{only} two PDSs that show clear peaks above the  99.9\% confidence limit, as described below. However, we would like to note that while the PDSs are consistent with power law noise with $\beta\sim$ $-$1.5--2.5 with a mean value of $-$2.0, there are pointings that exhibit more structure in the PDS around 0.01 Hz with multiple peaks just below the 99.9\% significance level. 


To identify the state of the system when these PDSs arise, we calculated the fraction of data as a function of phase whose root-mean square values are above the level of a standard deviation $+$ the median rms value of all the data (see Fig. \ref{fig-1}, black histogram). In addition, each fraction was further divided among different radio/X-ray states (coloured histograms in Fig. \ref{fig-1}). Quiescent radio/X-ray states are not shown in Fig. \ref{fig-1} as they have lower rms values. Fig. \ref{fig-1} shows that most of the rms variation is due to the rms-flux relation (tracing the orbital modulation), but two anomalous peaks show up at phases 0.3--0.4 and 0.6--0.7. The first peak is due to only the FHXR state and the second peak results from the excesses in the FIM/FSXR states. These states and related phases are the same states and phases where the QPOs discussed in this work are found.

Fig. \ref{fig-2} and Table \ref{table-2} show the PDSs and significances (based on the Monte-Carlo analysis referred to in Section 2) respectively of the QPOs we have identified in this subset of \rxte\/ data with Fig. \ref{fig-3} highlighting the data from the 2000 Apr 3 observation. As noted in Table \ref{table-2}, the QPOs occurred during the flaring states -- particularly in the FSXR/FHXR states. In the context of phase, we note that the QPOs occurred between phase 0.2--0.7. Interestingly, the QPOs discussed in \citet{vanderklis} all appear exclusively in the phase interval 0.0--0.75 that corresponds to the rising branch and top of the phase-folded X-ray lightcurve \citep{vilhu1}.

\subsubsection{2000 Apr 3}

This pointing occurred 2.2 days after the peak of a major flare ($\sim$13 Jy in the 15 GHz band) when the source was in the FSXR state. We note the presence of one 8.5 mHz ($\sim$ 120 s) QPO in this pointing, occurring in two separate episodes each lasting about 12 cycles (data between these two separate episodes are devoid of QPOs). In the PDS this QPO corresponds to 3\% rms and most of its power is concentrated in the three cycles visible at the beginning of the lightcurve in Fig. \ref{fig-2}\textit{b}. A blow-up of this region is plotted in Fig. \ref{fig-3} where there is possible indication of a weaker 30 mHz ($\sim$ 30 s) QPO-like period (seen in the PDS as a multiple-peaked region but not producing a peak above the 99.9\% significance level) and 100 mHz ($\sim$ 10 sec) QPO-like period (not seen in the PDS, however possibly contributing power only in this short segment of lightcurve) in alternating rhythm: three times the 30 sec period which form the peak of the 120 sec QPO and four times the 10 sec period which form the valley of the 120 sec QPO. 

\subsubsection{2009 Aug 9}

This pointing occurred 17 days after a major flare event ($\sim$ 1.5 Jy in the 15 GHz band). During this pointing the 15 GHz radio flux density is $\sim$ 50 mJy and the source is in the FHXR state. There is a strong 21 mHz ($\sim$ 50 s) QPO lasting about 20 cycles corresponding to 2\% rms in the PDS. In the PDS there are also two peaks with lower significances that correspond to frequencies 9 and 31 mHz.

\begin{table*}
\centering
\caption{Log of QPOs identified from Cyg X-3 in this study.}
\begin{tabular}{ccccccc}
\hline
ObsID & Date & MJD & Phase & TSMF$^{a}$ (days) & X-ray state$^{b}$ & obs. length (s) \\
\hline
50062 & 2000 Apr 3 & 51637.53--51637.62 & 0.33--0.78 & 2.16 & FSXR & 4526.6 \\
94328 & 2009 Aug 9 & 55052.82--55052.84 & 0.22--0.34 & 17.0 & FHXR & 1974.0 \\
\hline
\hline
  & no. PDS in ave. & Model $\beta$ & Model $\chi^{2}$/d.o.f. & Centroid freq. (mHz) & Significance \\
\hline
50062 &  3 & -2.0 & 41.7/27 & 8.5/30 & $>$99.9\%/$\sim$99\% \\  
94328 &  2 & -1.8 & 22.5/27 & 9/21/31 & $\sim$99\%/$>$99.9\%/$\sim$99.9\% \\
\hline
\hline
\label{table-2}
\end{tabular}
\begin{list}{}{}
\item[$^{\mathrm{a}}$] Time Since Major Flare. 
\item[$^{\mathrm{b}}$] Classification of K10 where FSXR/FHXR refer to the ``high'' state in the canonical nomenclature.
\end{list}
\end{table*}

\section[]{Discussion}

An in-depth analysis is beyond the scope of this paper. Here we speculate briefly on the origin of the QPOs found in this study. \\

$\bullet$ {\it QPOs and the {\bf geometry} of the system.} If we assume that the QPOs are X-ray emitting blobs or regions in orbit around the compact object, the possible timescales are determined by the innermost stable circular orbit (ISCO, highest frequency) and the Roche Lobe radius or the bow shock that is formed as the compact object plows the stellar wind of the companion (lowest frequency). Identifying the Keplerian radius of the QPOs ($\sim$0.16R$_{\odot}$) with the ISCO results in a compact object of thousands of solar masses therefore rendering this scenario rather unlikely. At the other end of the length scale, we find the Roche Lobe radius to be approximately 0.7--2.1 R$_{\odot}$ and the characteristic length scale of the bow shock, $R_{BS} = \dot{m} v_{wind} / 4 \pi \rho V^{2}_{C}\sim$ 3.1--5.7 R$_{\odot}$, with the compact object moving with velocity V$_{C}$ through the stellar wind of density $\rho$. The above values have been calculated using the mass ratio of the system (M$_{WR}$/M$_{C}$ = 3.8$^{+1.7}_{-1.4}$ given in \citealt{vilhu2}), orbital period of 4.8 hours, compact object mass M$_{C}$ = 1.4--30 M$_{\odot}$ and typical parameters for wind accretion from WR stars ($v_{wind}$ = 1600 km s$^{-1}$ with velocity law following \citealt{springmann} and mass-loss rate $\dot{m}$ = 1.6 $\times$ 10$^{-5} M_{\odot}$ yr$^{-1}$ following \citealt{szostek}). Therefore, for a reasonable range of compact object masses the QPOs seem not to be associated with these radii. \\  

$\bullet$ {\it QPOs as {\bf disk} oscillations?} In the literature there are several methods as to how low-frequency QPOs can be invoked in the accretion disk, e.g.  low-frequency ``dynamo cycles" in the azimuthal magnetic field \citep{oneill} or Accretion-Ejection Instability \citep{varniere} to name a few. However, the geometry of the system might present a challenge to disk-based phenomena. If the accretion disk is perpendicular to the jet and if the jet is inclined $\sim$ 14 degrees to the line of sight \citep{mio}, then we should be viewing the disk face-on. The problem then arises as to how the actual modulation can be seen at all. A further interesting question is how would the simultaneous frequencies observed in the 2000 Apr 3 pointing arise in the disk, especially when they seem to be producing a clear pattern? Are they causally linked and/or produced by the same underlying mechanism? 
Interestingly, the 2009 Aug 9 pointing indicates that the peaks in the PDS form a 3:2:1 frequency relation, the same relation as has been reported from Sgr A* \citep{aschenbach}. \\

$\bullet$ {\it QPOs as {\bf coronal} oscillations?}
As the QPOs appear to be present in energy bands from 2--15 keV and as they appear to arise in flaring radio/X-ray states that can be modelled solely with a Comptonization component (K10) this could imply that the actual component causing the Comptonization of soft photons is oscillating. An oscillating corona due to a magneto-acoustic wave propagating within the corona producing multiple QPOs has been theoretically discussed in \citet{cabanac}. \\

$\bullet$ {\it QPOs as oscillations due to the {\bf jet}?}
As the QPOs appear to occur after major radio flares, the latter most likely signaling a jet ejection event, the two are probably linked. This connection could be two-fold: either the jet is shadowing, therefore modulating, an underlying oscillation in or emission from the disk or corona, or the QPOs are caused by some structure in the jet (emitting X-rays, e.g. a shock). The latter scenario is discussed for blazars in \citet{rani} and it could occur in a turbulent region behind a shock where dominant eddies with turnover times corresponding to the QPO periods could explain the short-lived, quasi-periodic fluctuation.  It is also possible that the jet possesses a helical structure, and a relativistic shock propagating down the jet encounters regions of enhanced magnetic field and/or electron density intensifying the emission and causing fluctuations.\\   

$\bullet$ {\it QPOs as oscillations due to the {\bf wind}?}
Similarly to the jet case, clumps in the Wolf-Rayet companion's stellar wind could cause a shadowing effect on an underlying oscillation/emission. The pointings that exhibit QPOs are often accompanied by sudden drops in intensity (\citealt{vanderklis}; their 1983 Nov 5 observation). A quite similar 15\% drop in the lightcurve can be seen in Fig. \ref{fig-2}\textit{a}. The light crossing time in this drop is about 16 sec and implies an upper limit on the size of the emission region radius of $<2.4\times10^{11}$ cm. This value is remarkably close to the separation between the components $a \simeq 3\times 10^{11} (M/30M_{\odot})^{1/3}$ cm,  and thus could be produced by a light-blocking clump. A clumpy wind was established in \citet{szostek}.
	 
\section{Summary}

In this paper, we reported on the detection of QPOs in Cygnus X-3. In this initial study, we found two occasions in the RXTE data where QPOs were present, and these followed major radio flares. We speculated on various scenarios that may explain the QPOs, but a more in-depth study is necessary to single out a favoured one.



\section*{Acknowledgments}

The authors thank the referee for the careful reading of the manuscript and excellent suggestions that significantly improved the paper. KIIK acknowledges the Finnish Graduate School in Astronomy and Space Physics. The authors gratefully thank the Vilho, Yrj\"o and Kalle V\"ais\"al\"a Foundation for a travel grant that stimulated this discussion. This research has made use of data obtained from the High Energy Astrophysics Science Archive Research center (HEASARC), provided by NASA's Goddard Space Flight Center.


\begin{thebibliography}{99}  

\bibitem[\protect\citeauthoryear{Aschenbach et al.}{2004}]{aschenbach} Aschenbach, B., Grosso, N., Porquet, D., Predehl, P. 2004, A\&A, 417, 71

\bibitem[\protect\citeauthoryear{Axelsson, Larsson \& Hjalmarsdotter}{Axelsson et al.}{2009}]{axelsson} Axelsson M., Larsson S., Hjalmarsdotter L., 2009, MNRAS, 394, 1544

\bibitem[\protect\citeauthoryear{Benlloch et al.}{2001}]{benlloch} Benlloch S., Wilms J., Edelson R., Yaqoob T., Staubert R., 2001, ApJ, 562, L121

\bibitem[\protect\citeauthoryear{Berger \& van der Klis}{1994}]{berger} Berger M., van der Klis M., 1994, A\&A, 292, 175  

\bibitem[\protect\citeauthoryear{Cabanac et al.}{2010}]{cabanac} Cabanac, C., Henri, G., Petrucci, P.-O., Malzac, J., Ferreira, J.,  Belloni, T. M., 2010, MNRAS, 404, 738

\bibitem[\protect\citeauthoryear{Choudhury et al.}{2004}]{choudhury} Choudhury M., Rao A. R., Vadawale S. V., Jain A. K., Singh N. S., 2004, A\&A, 420, 665  

\bibitem[\protect\citeauthoryear{Dickey \& Lockman}{1990}]{dickey} Dickey J. M., Lockman F. J., 1990, ARA\&A, 28, 215  

\bibitem[\protect\citeauthoryear{Giacconi et al.}{1967}]{giacconi} Giacconi R., Gorenstein P., Gursky H., Waters J. R., 1967, ApJ, 148, L119   

\bibitem[\protect\citeauthoryear{Houck \& Denicola}{2002}]{houck} Houck J. C., Denicola L. A., 2000, ASPC, 216, 591  

\bibitem[\protect\citeauthoryear{Koljonen et al.}{2010}]{koljonen} Koljonen K. I. I., Hannikainen D. C., McCollough M. L., Pooley G. G., Trushkin S. A., 2010, MNRAS, 406, 307 (K10)  

\bibitem[\protect\citeauthoryear{Koljonen et al.}{2011}]{koljonen2} Koljonen K. I. I., Hannikainen D. C., McCollough M. L., Pooley G. G., Trushkin S. A., Tavani M., Droulans R., 2011, IAUS, 275, 285K 

\bibitem[\protect\citeauthoryear{Miller-Jones et al.}{2004}]{millerjones} Miller-Jones J. C. A., Blundell K. M., Rupen M. P., Mioduszewski A. J., Duffy P., Beasley A. J., 2004, ApJ, 600, 368

\bibitem[\protect\citeauthoryear{Mioduszewski et al.}{2001}]{mio} Mioduszewski A. J., Rupen M. P., Hjellming R. M. et al., 2001, ApJ, 553, 766  

\bibitem[\protect\citeauthoryear{O{'}Neill et al.}{2010}]{oneill} O'Neill S. M., Reynolds C. S., Miller M. C., Sorathia K. A., 2010, submitted to ApJ

\bibitem[\protect\citeauthoryear{Parsignault et al.}{1972}]{parsignault} Parsignault D. R., Gursky H., Kellogg E. M. et al., 1972, Nat. Phys. Sci., 239, 123  

\bibitem[\protect\citeauthoryear{Predehl et al.}{2000}]{predehl} Predehl P., Burwitz V., Paerels F., Tr\"umper J., 2000, A\&A, 357, L25  

\bibitem[\protect\citeauthoryear{Rani et al.}{2010}]{rani} Rani B., Gupta A. C., Joshi U. C., Ganesh S., Wiita P. J., 1991, ApJ, 719, L153  

\bibitem[\protect\citeauthoryear{Rao, Agrawal \& Manchanda}{Rao et al.}{1991}]{rao} Rao A. R., Agrawal P. C., Manchanda R. K., 1991, A\&A , 241, 127  

\bibitem[\protect\citeauthoryear{Scott et al.}{2003}]{scott} Scott, D.~M., Finger, M.~H., \& Wilson, C.~A., 2003, MNRAS, 344, 412

\bibitem[\protect\citeauthoryear{Singh et al.}{2002}]{singh} Singh N. S., Naik S., Paul B., Agrawal P. C., Rao A. R., Singh K. Y., 2002, A\&A, 392, 161

\bibitem[\protect\citeauthoryear{Szostek \& Zdziarski}{2008}]{szostek} Szostek A., Zdziarski A. A., 2008, MNRAS, 386, 593

\bibitem[\protect\citeauthoryear{Springmann}{1994}]{springmann} Springmann U., 1994, A\&A, 289, 505

\bibitem[\protect\citeauthoryear{Timmer \& K\"onig}{1995}]{timmer} Timmer J., K\"onig M., 1995, A\&A, 300, 707

\bibitem[\protect\citeauthoryear{van der Klis \& Jansen}{1985}]{vanderklis} van der Klis M. \& Jansen F. A., 1985, Nat, 313, 768  

\bibitem[\protect\citeauthoryear{van Kerkwijk et al.}{1992}]{keerkwijk} van Kerkwijk M. H., Charles P. A., Geballe T. R. et al., 1992, Nat, 355, 703  

\bibitem[\protect\citeauthoryear{Waltman et al.}{1996}]{waltman} Waltman E. B., Foster R. S., Pooley G. G., Fender R. P., Ghigo F. D., 1996, AJ, 112, 2690

\bibitem[\protect\citeauthoryear{Varni\`ere \& Tagger}{2002}]{varniere} Varni\`ere P., Tagger M., 2002, A\&A, 394, 329

\bibitem[\protect\citeauthoryear{Willingale, King \& Pounds}{Willingale et al.}{1985}]{willingale} Willingale R., King A. R., Pounds K. A., 1985, MNRAS, 215, 295

\bibitem[\protect\citeauthoryear{Vilhu et al.}{2003}]{vilhu1} Vilhu O., Hjalmarsdotter L., Zdziarski A. A. et al., 2003, A\&A, 411, L405 

\bibitem[\protect\citeauthoryear{Vilhu et al.}{2009}]{vilhu2} Vilhu O., Hakala P., Hannikainen D., McCollough M., Koljonen K., 2009, A\&A, 501, 679

\bibitem[\protect\citeauthoryear{Zdziarski, Misra \& Gierli\'nski}{Zdziarski et al.}{2010}]{zdziarski} Zdziarski A. A., Misra R., Gierli\'nski M., 2010, MNRAS, 402, 767. 

\end{thebibliography}
\end{document}